\documentclass{article}

\usepackage{PRIMEarxiv}

\usepackage[utf8]{inputenc} 
\usepackage[T1]{fontenc}    
\usepackage{hyperref}       
\usepackage{url}            
\usepackage{booktabs}       
\usepackage{amsfonts}       
\usepackage{nicefrac}       
\usepackage{microtype}      
\usepackage{lipsum}
\usepackage{fancyhdr}       
\usepackage{graphicx}       
\graphicspath{{media/}}     
 \usepackage{multirow}
\usepackage{amsmath}
\usepackage{booktabs}
\usepackage{graphicx}
\usepackage{subcaption}
\usepackage{algorithm}
\usepackage{algorithmic}
\usepackage{xcolor}
\usepackage{bm}
\usepackage{amssymb}

\title{A Privacy-Preserving UAV-Assisted Secure Wireless Communication and Computing System Based on Fully Homomorphic Encryption}

\author{Quanxi~Zhou$^{1}$,
Wencan~Mao$^{2}$, Ye~Tao$^{1}$,~Manabu~Tsukada$^{1}$,~Kensuke~Fukuda$^{2}$,~and~Yusheng~Ji$^{2}$\\
\thanks{The corresponding auther is Quanxi~Zhou~(usainzhou@g.ecc.u-tokyo.ac.jp).},
\thanks{This work was supported by the JST ASPIRE (JPMJAP2325), JSPS KAKENHI Grant No. JP24K02937, and JST SPRING GX (JPMJSP2108).},
$^{1}$The University of Tokyo, Tokyo, Japan\\
$^{2}$National Institute of Informatics, Tokyo, Japan\\
\{usainzhou, tydus, mtsukada\}@g.ecc.u-tokyo.ac.jp, \{wencan\_mao, kensuke, kei\}@nii.ac.jp.
}

\begin{document}
\maketitle
\begin{abstract}
With the advancement of beyond-5G networks, privacy-preserving communication and secure computation face critical challenges due to the openness of wireless channels. Existing security mechanisms protect data transmission but fail to secure the computation process itself. In this paper, we propose a UAV-assisted secure computing framework that integrates Cheon Kim Kim Song (CKKS) fully homomorphic encryption (FHE) with multi-party secure computation (MPC). Leveraging UAV flexibility, our framework decouples key distribution from encrypted computation, enabling secure encrypted weighted aggregation (EWA), encrypted matrix multiplication (EMM), and encrypted decision tree (EDT) inference. Experimental results show that our framework outperforms existing schemes: for EDT, it supports floating-point operations and achieves better time efficiency than Brakerski-Fan-Vercauteren (BFV); for EMM, it demonstrates superior time and space efficiency compared to Brakerski-Gentry-Vaikuntanathan (BGV) and Somewhat Homomorphic Encryption (SHE); \textcolor{black}{while for EWA, it ensures high numerical precision and robustness across varying data length}.
\end{abstract}

\keywords{
Unmanned Aerial Vehicle, Fully Homomorphic Encryption, Multi-Party Secure Computation, CKKS.}

\section{Introduction}

\begin{figure}[t]
    \centering
    \includegraphics[width=0.95\textwidth, keepaspectratio]{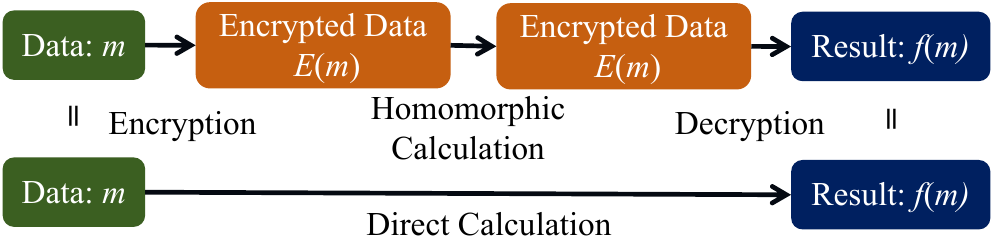}
    \caption{The basic principles and process of FHE.}
\label{fig: fhe}
    \setlength{\belowcaptionskip}{-10pt}  
\end{figure}

The rapid evolution of beyond-5G communication technologies 
has significantly increased the demand for privacy-preserving data processing in wireless networks~\cite{privacy}.
Due to the openness, broadcast nature, and dynamicity of wireless channels, both parties involved in data exchange face significant security challenges~\cite{channel_challenge}.
cryptographic, physical layer, and protocol layer approaches.
While the aforementioned methods can ensure communication security, they cannot guarantee the security of the data computation process. As Fig.\ref{fig: fhe} shows, fully homomorphic encryption (FHE) technology, a cryptographic primitive that allows arbitrary computations to be carried out directly on ciphertexts, producing an encrypted result which, when decrypted, matches the outcome of the same operations applied to the plaintext, can address this issue~\cite{intro_fhe}. The security of FHE is guaranteed by the Ring Learning With Errors (RLWE) problem~\cite{rlwe}, whose computational hardness ensures that encrypted data remains confidential even during arbitrary computations.
Despite its strong cryptographic guarantees, practical deployment of FHE-based systems faces two major challenges: high computational overhead and the requirement of a reliable computation intermediary.

To address the above challenges and meet the demands of \textbf{personal privacy computation} services such as reputation evaluation, health assessment, and resource matching, this paper aims to design a multi-party secure computation (MPC) system that supports both encrypted computation and secure communication.
The proposed framework enables two data owners to perform joint computations and exchange results without revealing their plaintext data to each other. To mitigate the high computational overhead of FHE, we adopt the Cheon Kim Kim Song (CKKS) scheme~\cite{CKKS}, which offers superior efficiency for real-number operations. Based on CKKS, we design ciphertext computation algorithms, including encrypted weighted aggregation (EWA), encrypted matrix multiplication (EMM), and encrypted decision tree (EDT), to support diverse application requirements.
Furthermore, to ensure that both parties operate exclusively on encrypted data and to mitigate the risks of both malicious dishonesty and honest-but-curious behavior (where the intermediary attempts to infer as much information as possible) in third-party computation relays, we employ two unmanned aerial vehicles (UAVs) as FHE computation intermediaries. Leveraging the flexibility, mobility, and low cost of UAVs, the system delegates key distribution, encrypted computation, and authentication to aerial relays, ensuring end-to-end data confidentiality even under dishonest participants or UAV eavesdropping.

The main contributions are summarized as follows:
\begin{itemize}
\item \textbf{We propose a UAV-assisted privacy-preserving computing framework} that integrates CKKS-based FHE with MPC mechanism. The framework enables two data owners to perform secure collaborative processing in wireless environments, addressing both communication security and computation privacy challenges through a novel UAV-relayed architecture.

\item \textbf{Within this framework, we design and implement three core encrypted computation modules}: EWA, EMM, and EDT, to support diverse application requirements. By spanning simple arithmetic, matrix operations, and complex machine learning, these modules collectively demonstrate the \textbf{versatility} of our protocol. We provide formal analysis of their correctness and security properties under both malicious and honest-but-curious adversary models.

\item \textbf{We conduct extensive simulations in realistic UAV-assisted network scenarios} to evaluate computational accuracy, time efficiency, and memory consumption. Comparative analysis with the \textbf{BFV}~\cite{sortinghat}, \textbf{SHE}~\cite{SHE}, and \textbf{BGV-based}~\cite{BGV} approaches demonstrates that our framework outperforms existing solutions while maintaining strong security guarantees, thereby validating its practical feasibility for real-world deployment.
\end{itemize}

\section{Related Works}
Privacy-preserving computation (PPC) is playing an increasingly important role in secure computing and communication~\cite{importance}. 
As a crucial PPC technology, FHE offers several advantages, including low communication overhead, minimal interaction requirements, support for outsourced computation, and strong security guarantees for algorithms.
However, existing encryption schemes such as SHE~\cite{SHE} and BGV~\cite{BGV}, as well as the BFV~\cite{sortinghat} for encrypted data classification, suffer from long computation times and high complexity. With the introduction of the CKKS~\cite{CKKS} and the application of OpenFHE~\cite{openfhe}, the new generation of FHE achieves significantly improved computational performance.


Furthermore, the use of UAVs as intermediate computation relays for fully FHE and PPC is emerging as a promising research area. In~\cite{uav-1}, a framework integrating attribute-based encryption with FHE enables efficient processing of aggregated encrypted data by UAVs. In~\cite{uav-2}, a homomorphic encryption-based distributed estimation protocol reduces computational overhead under eavesdropping attacks. In~\cite{uav-3}, a secure navigation framework combining reinforcement learning with FHE supports real-time processing of encrypted UAV video streams while reducing latency.
\section{System Design}

We propose a new UAV-assisted secure computation and communication system.
Based on MPC, a four-party collaborative model involving Alice $P_A$, Bob $P_B$, UAV-1 $P_{U-1}$, and UAV-2 $P_{U-2}$ is designed in this system. Within this framework, the computation requester, Alice, and the computation participant, Bob, provide their respective encrypted data and pre-agreed verification ciphertexts $x_A$ and $x_B$, which can ensure the correctness of the computation. \textcolor{black}{Meanwhile, UAV-1 and UAV-2 perform distinct tasks as follows to achieve the separation of key and data.
The key generator, UAV-1, provides the key group $x_{U-1}$, including public key $pk$, secret key $sk$, and evaluation key $evk$, which enables the server to perform specific operations on ciphertexts without exposing the secret key.
The computation service provider, UAV-2, supplies FHE-based encrypted data operation $x_{U-2}$ and secure computational services.}
We define the secure protocol as $\pi$, the MPC process is represented as
\begin{equation}
    f_m(x_A, x_B, x_{U-1}, x_{U-2}) \leftarrow \pi f_m(x_A, x_B, x_{U-1}, x_{U-2})
\end{equation}

The joint output $y_m$ can be represented as
\begin{equation}
    y_m = f_m(x_A, x_B, x_{U-1}, x_{U-2})
\end{equation}

\begin{figure}[t]
    \centering
    \includegraphics[width=0.85\textwidth, keepaspectratio]{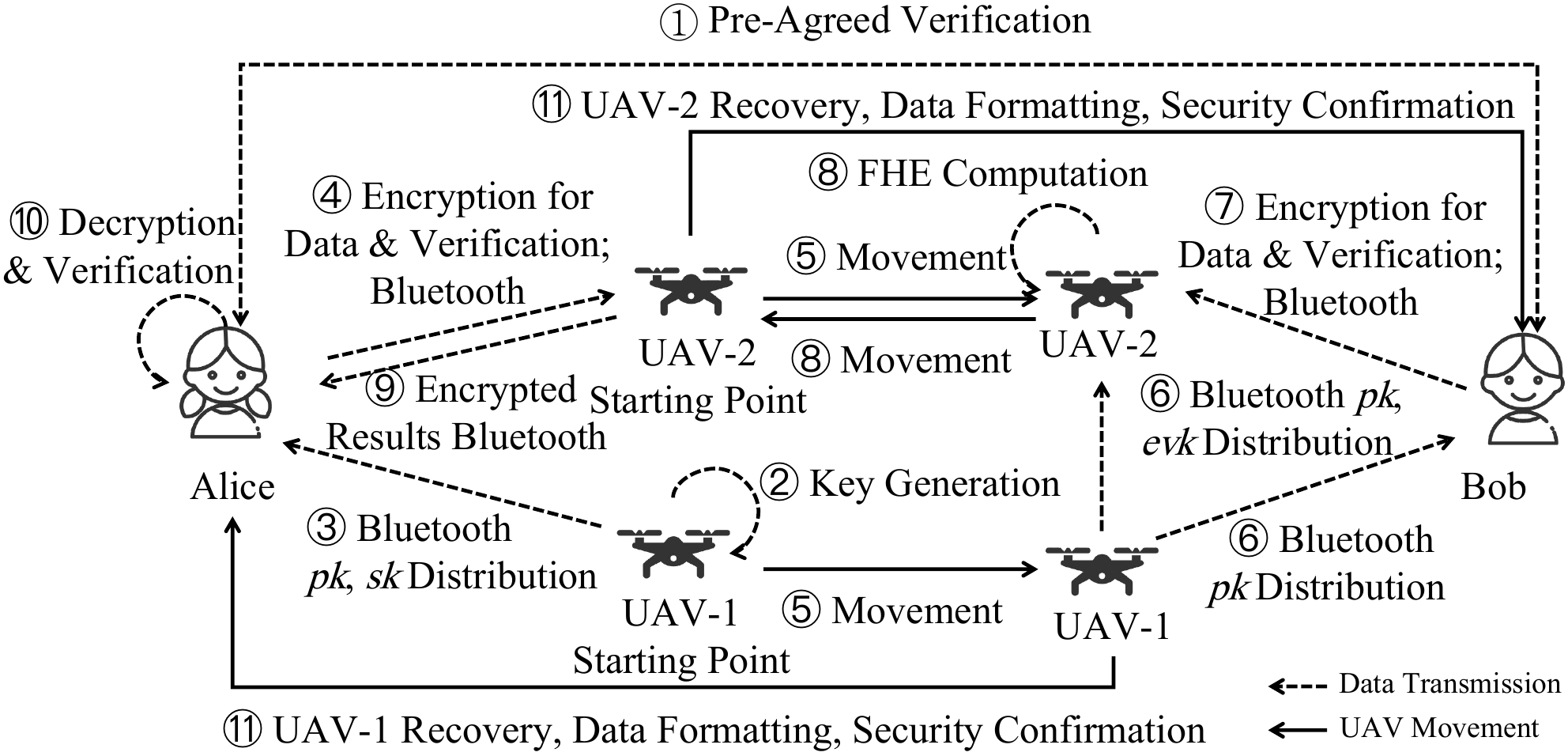}
    \caption{The protocol of UAV-assisted FHE secure computation and communication.}
\label{fig: protocol}
    \setlength{\belowcaptionskip}{-10pt}  
\end{figure}

As shown in Fig.~\ref{fig: protocol}, the secure protocol $\pi$ is described as
\textcolor{black}{
\begin{enumerate}
    \item[1] Alice and Bob jointly determine the verification.
    \item[2] UAV-1 executes $\{pk,sk,evk\} \leftarrow \text{KeyGen}(params)$ to generate the cryptographic keys.
    \item[3] Through Bluetooth, UAV-1 distributes $\{pk,sk\}$ to Alice. 
    \item[4] Alice encrypts data and sends the ciphertext to UAV-2.
    \item[5] UAV-1 and UAV-2 choose different routes and move toward Bob, potentially involving trajectory optimization.
    \item[6] UAV-1 distributes $pk$ to Bob and $\{pk,evk\}$ to UAV-2 through Bluetooth.
    \item[7] Bob encrypts data and transmits the ciphertext and verification to UAV-2.
    \item[8] UAV-2 flies towards Alice and processes computation.
    \item[9] UAV-2 transmits the result ciphertext to Alice.
    \item[10] Alice decrypts the ciphertext to obtain the results and confirms the verification. 
    \item[11] UAV-1 returns to Alice and UAV-2 returns to Bob. They operate security confirmation and data formatting. 
\end{enumerate}}

\textcolor{black}{Note that Alice, Bob, UAV-1, and UAV-2 follow the honest majority assumption (i.e., the semi-honest parties, who execute the protocol cycle but tend to learn as much as possible,  exceed half of the total parties).}
   
\textbf{Security Analysis}

\begin{itemize}

    \item Data Confidentiality:
    Alice and Bob encrypt their private inputs using CKKS $c = \text{Enc}_{pk}(x)$, where $\text{Enc}(\cdot)$ represents encryption process.
    Under the RLWE assumption, the scheme achieves indistinguishability under chosen plaintext attack (IND-CPA) security.
    Therefore, without the secret key $sk$, neither UAV-2 nor any eavesdropper
    can recover the plaintext data.

    \item Secret Key Protection:
    The secret key $sk$ is never revealed
    to Bob or UAV-2, which cannot decrypt intermediate or final
    ciphertexts. And by confirmation process, Alice can ensure UAV-1 is not intercepted. 

    \item Secure Computation:
    UAV-2 performs homomorphic evaluation $c_f = \text{Eval}_{f,evk}(c_A, c_B)$,
    where $\text{Eval}_{f,evk}(\cdot)$ represents the FHE
    algorithm that computes a function over ciphertexts.
    Since all operations are conducted
    over ciphertexts, data privacy is preserved.

    \item Integrity Verification:
    The jointly determined verification mechanism enables Alice to confirm the correctness of the returned ciphertext and UAV-2 is not intercepted. 
    
    \item Data Formatting: After computation, Alice and Bob physically recover UAV-1 and UAV-2 and securely format all onboard data to prevent residual information leakage.

\end{itemize}

Therefore, assuming the hardness of the RLWE problem and secure key
distribution, the proposed protocol preserves data confidentiality and
computation correctness.

\section{Algorithm}

\subsection{CKKS-based Encrypted Weighted Aggregation}
We define the standard weight aggregation (WA) as
\begin{equation}
S = \sum_{i=1}^{n} x_i w_i ,
\end{equation}
where $x = \{x_1,\ldots,x_n\} \in \mathbb{R}^n$ represents the data vector
and $w = \{w_1,\ldots,w_n\} \in \mathbb{R}^n$ represents the weight vector.

First, CKKS parameters are generated by 
$\text{ParamsGen}(1^\lambda)$, producing public parameters $params$,  where $\lambda$ represents security parameter. 
Then, key pair $(sk, pk)$ is generated by 
$\text{KeyGen}(params)$, followed by generation of evaluation keys 
$evk$ (including relinearization $rlk$ and rotation keys $rotk$).

The vectors $x$ and $w$ are encoded and encrypted as
\begin{equation}
c_x = \text{Enc}_{pk}(x), \quad
c_w = \text{Enc}_{pk}(w).
\end{equation}

The encrypted weighted sum is evaluated through slot-wise
homomorphic multiplication by relinearization and rescaling:
\begin{equation}
c = \text{Rescale}(\text{Relin}(c_x \otimes c_w)).
\end{equation}
where, $\text{Relin}(\cdot)$ denotes the relinearization operation, and $\text{Rescale}(\cdot)$ represents the CKKS rescaling operation.
Then, slot aggregation is performed to sum all packed elements:
\begin{equation}
c_S = \text{Aggregate}(c),
\end{equation}
where $\text{Aggregate}(\cdot)$ represents slot-wise summation
implemented through ciphertext rotations $\text{Rot}(\cdot)$
and homomorphic additions. The rotation operation
$\text{Rot}(\cdot)$ shifts the packed slots using the
rotation keys $rotk$ to enable accumulation across slots. And the detailed process can be expressed as \textbf{Algorithm~\ref{alg:ckks_weighted_sum}}.


\begin{algorithm}[h]
\caption{CKKS-Based Encrypted Weighted Aggregation}
\label{alg:ckks_weighted_sum}
\begin{algorithmic}[1]
\REQUIRE Data vector $x \in \mathbb{R}^n$, weight vector $w \in \mathbb{R}^n$
\ENSURE Weighted sum $S = \sum_{i=1}^{n} w_i x_i$

\STATE Generate CKKS parameters via $\text{ParamsGen}(1^\lambda)$
\STATE Generate keys $(sk, pk, evk)$, where $evk = \{rlk, rotk\}$
\STATE $c_x \leftarrow \text{Enc}_{pk}(\text{Encode}(x))$; $c_w \leftarrow \text{Enc}_{pk}(\text{Encode}(w))$

\STATE $c \leftarrow c_x \otimes c_w$; $c \leftarrow \text{Relin}(c, rlk)$; $c \leftarrow \text{Rescale}(c)$

\FOR{$i=0$ to $\log_2(n)-1$}
\STATE $c \leftarrow c + \text{Rot}(c,2^i,rotk)$
\ENDFOR

\STATE $c_S \leftarrow c$; $S \leftarrow \text{Dec}_{sk}(c_S)$

\RETURN $S$

\end{algorithmic}
\end{algorithm}

\textbf{Correctness Analysis}

Assume that $c_x = \text{Enc}_{pk}(x)$ and $c_w = \text{Enc}_{pk}(w)$.
By the homomorphic property of CKKS,
\begin{equation}
  \text{Dec}_{sk}(c_x \otimes c_w)
= x \odot w + e_1,  
\end{equation}
where $e_1$ represents the bounded approximation error.

After rescaling and slot aggregation, the resulting ciphertext $c_S$ can be represented as:
\begin{equation}
 \text{Dec}_{sk}(c_S)
= \sum_{i=1}^{n} x_i w_i + e,   
\end{equation}
where $e$ is determined by the scale factor and noise growth.

Under valid parameter selection, the accumulated error remains within the acceptable precision range. Therefore, the output correctly approximates the weighted sum.

\textbf{Security Analysis}

In the proposed CKKS-based encrypted weighted aggregation, only ciphertexts
$c_x$, $c_w$, and intermediate ciphertexts are processed. The plaintext vectors $x$ and $w$ are never exposed.
The CKKS scheme is IND-CPA secure under RLWE assumption. Thus, for any probabilistic polynomial-time adversary,
ciphertexts are computationally indistinguishable.
Since the secret key $sk$ is not involved in the evaluation phase, recovering plaintext information from ciphertexts is equivalent to solving the RLWE problem. Hence, the proposed encrypted weighted summation preserves data confidentiality.

\subsection{Encrypted Matrix Multiplication}
We define the matrix multiplication (MM) process as
\begin{equation}
C = AB,
\end{equation}
where
$A \in \mathbb{R}^{m \times k}$,
$B \in \mathbb{R}^{k \times n}$,
and
$C \in \mathbb{R}^{m \times n}$.


The matrices are first encoded into CKKS plaintext slots and then encrypted as
\begin{equation}
c_A = \text{Enc}_{pk}(\text{Encode}(A)), \quad
c_B = \text{Enc}_{pk}(\text{Encode}(B)).
\end{equation}

The encrypted MM is evaluated by homomorphic
ciphertext multiplications together with slot rotations and additions:
\begin{equation}
c_C = \text{MatMul}(c_A, c_B),
\end{equation}
where $\text{MatMul}(\cdot)$ denotes homomorphic MM procedure implemented through ciphertext multiplications followed by relinearization $\text{Relin}(\cdot)$, rescaling $\text{Rescale}(\cdot)$,
and slot rotations $\text{Rot}(\cdot)$.
Plaintext result is represented as
\begin{equation}
C = \text{Dec}_{sk}(c_C).
\end{equation}
The detailed computation procedure is shown in
\textbf{Algorithm~\ref{alg:ckks_matrix_mul}}.
\begin{algorithm}[h]
\caption{CKKS-Based Encrypted Matrix Multiplication}
\label{alg:ckks_matrix_mul}
\begin{algorithmic}[1]
\REQUIRE $A \in \mathbb{R}^{m \times k}$,
         $B \in \mathbb{R}^{k \times n}$
\ENSURE $C = AB$

\STATE Generate CKKS parameters
\STATE Generate keys $(sk, pk, evk)$, where $evk =\{rlk, rotk\}$

\STATE Encode matrices:
      $m_A \leftarrow \text{Encode}(A)$
      $m_B \leftarrow \text{Encode}(B)$

\STATE Encrypt:
      $c_A \leftarrow \text{Enc}_{pk}(m_A)$
      $c_B \leftarrow \text{Enc}_{pk}(m_B)$

\FOR{each column $j$ of $B$}
    \STATE $c \leftarrow c_A \otimes \text{Rot}(c_B,j)$
    \STATE $c \leftarrow \text{Relin}(c, rlk)$
    \STATE $c \leftarrow \text{Rescale}(c)$
    \STATE Accumulate partial sums
\ENDFOR

\STATE $C \leftarrow \text{Dec}_{sk}(c_C)$
\RETURN $C$
\end{algorithmic}
\end{algorithm}

\textbf{Correctness Analysis}

For WA, by the homomorphic property of CKKS,
\begin{equation}
\text{Dec}_{sk}(c_x \otimes c_w)
= x \odot w + e_1,
\end{equation}
where $e_1$ is bounded approximation error.
After aggregation,
\begin{equation}
\text{Dec}_{sk}(c_S)
=\sum_{i=1}^{n} x_i w_i + e.
\end{equation}
where $e$ represents the accumulated bounded error.
For matrix multiplication, each entry satisfies
\begin{equation}
\text{Dec}_{sk}(c_C)
=
AB + E,
\end{equation}
where $E$ represents accumulated approximation error.

Under proper parameter selection, the error remains within acceptable precision.
Thus, both operations correctly approximate their plaintext counterparts.

\textbf{Security Analysis}

Only ciphertexts $c_x$, $c_w$, $c_A$, $c_B$, and intermediate
ciphertexts are processed during evaluation, while plaintext
data are never exposed. Since the computation is performed
entirely in the encrypted domain and does not involve the
secret key $sk$, data confidentiality is preserved under
the security guarantees of the CKKS scheme.

\subsection{CKKS-based Encrypted Decision Tree Inference}

We define a binary decision tree (DT) with $T$ internal decision nodes.
For an input feature vector $x = \{x_1,\ldots,x_d\} \in \mathbb{R}^d$, each internal node $j$ is associated with a feature index $k_j$ and a
threshold $t_j$.
The node decision is defined as
\begin{equation}
b_j = \mathbf{1}(x_{k_j} - t_j \ge 0),
\end{equation}
where $\mathbf{1}(\cdot)$ denotes the indicator function.
The final prediction is obtained by evaluating the tree path:
\begin{equation}
y = f_T(b_1,\ldots,b_T),
\end{equation}
where $f_T(\cdot)$ represents the decision tree evaluation function.

To securely compute $y$, we adopt a leveled homomorphic encryption
scheme based on CKKS, which supports approximate arithmetic over
real numbers.

The feature vector is first encoded and then encrypted as
\begin{equation}
m_x = \text{Encode}(x), \quad
c_x = \text{Enc}_{pk}(m_x).
\end{equation}

For each internal node, the encrypted comparison is approximated by
first computing the difference
\begin{equation}
c_{d_j} = c_x[k_j] - t_j,
\end{equation}
where the subtraction is performed between a ciphertext component
and a plaintext constant.

Since direct comparison is not supported in CKKS, the indicator
function is approximated using a low-degree polynomial:
\begin{equation}
c_{b_j} = \text{PolyApprox}(c_{d_j}),
\end{equation}
where $\text{PolyApprox}(\cdot)$ represents a polynomial approximation
of the step function (e.g., sigmoid or Chebyshev approximation),
allowing comparison-like behavior in encrypted domain.

The encrypted prediction is then obtained through homomorphic
evaluation of the decision tree:
\begin{equation}
c_y = \text{EvaluateTree}(c_{b_1},\ldots,c_{b_T}),
\end{equation}
where $\text{EvaluateTree}(\cdot)$ denotes the encrypted tree
evaluation procedure implemented via homomorphic additions and
multiplications to compute the leaf prediction corresponding to
the decision path.

The prediction result can be represented as
\begin{equation}
y = \text{Dec}_{sk}(c_y).
\end{equation}

The computational complexity is linear in the number of
nodes $T$. The detailed procedure is shown in
\textbf{Algorithm~\ref{alg:ckks_tree}}.

\begin{algorithm}[h]
\caption{CKKS-Based Encrypted Decision Tree}
\label{alg:ckks_tree}
\begin{algorithmic}[1]
\REQUIRE Feature vector $x \in \mathbb{R}^d$
\ENSURE Prediction $y$

\STATE Generate CKKS parameters via $\text{ParamsGen}(1^\lambda)$
\STATE Generate keys $(sk, pk, evk)$

\STATE $m_x \leftarrow \text{Encode}(x)$; $c_x \leftarrow \text{Enc}_{pk}(m_x)$

\FOR{each internal node $j$}
    \STATE $c_{d_j} \leftarrow c_x[k_j] - t_j$; $c_{b_j} \leftarrow \text{PolyApprox}(c_{d_j})$
\ENDFOR

\STATE $c_y \leftarrow \text{EvaluateTree}(c_{b_1},\ldots,c_{b_T})$
\STATE $y \leftarrow \text{Dec}_{sk}(c_y)$

\RETURN $y$
\end{algorithmic}
\end{algorithm}

\textbf{Correctness Analysis}

Let $c_x = \text{Enc}_{pk}(x)$.
By homomorphic property of CKKS,
\begin{equation}
\text{Dec}_{sk}(c_{d_j})
= x_{k_j} - t_j + e_1,
\end{equation}
where $e_1$ represents the bounded approximation error.

After polynomial approximation,
\begin{equation}
\text{Dec}_{sk}(c_{b_j})
= b_j + e_2,
\end{equation}
where $e_2$ depends on approximation degree and scaling factor.

Since tree evaluation only involves holomorphic additions and multiplications, the final ciphertext satisfies
\begin{equation}
\text{Dec}_{sk}(c_y)
= f(b_1,\ldots,b_T) + e,
\end{equation}
where $e$ is the accumulated bounded error.

Under appropriate parameter selection, the total error remains within the acceptable precision range. Therefore, encrypted inference approximates the plaintext model output.

\textbf{Security Analysis}

During encrypted inference, only ciphertexts $c_x$, $c_{d_j}$,
$c_{b_j}$, and $c_y$ are processed, while the plaintext feature
vector $x$ is never exposed. Since the inference procedure operates entirely in the encrypted domain and does not involve the secret key $sk$, data confidentiality is preserved under the security guarantees of the CKKS scheme.

\section{Experiment Results}

\subsection{Experimental Setting}
This study is conducted within \textcolor{black}{a real-world commercial credit evaluation~\cite{ESG} setting} to ensure practical relevance and external validity. 
We selected the DJI Matrice 300 RTK and integrated an onboard computing module equipped with a CPU-based processing unit to execute key generation and FHE tasks. The CKKS scheme was implemented using the OpenFHE library.
The detailed system specifications of the UAV-board are presented in Table~\ref{tab:env}. The CKKS FHE parameters are presented in Table~\ref{tab:hyperparameters}.
\begin{table}[h]
\centering
\caption{System Setting.}
\label{tab:env}
\begin{tabular}{l|l|l|l}
\toprule
\textbf{Category} & \textbf{Specification} &\textbf{Category} & \textbf{Specification} \\
\midrule
Operating System & Ubuntu 20.04 &
Bandwidth & 1~Mbps \\
CPU & 2~vCPU~Cores &
Memory & 4~GiB \\
Compiler & GCC 11.0 &
CMake Version & 3.14.0 \\
Applied Library & OpenFHE &
Database & MySQL 8.0\\
\bottomrule
\end{tabular}
\end{table}
\begin{table}[h]
\centering
\caption{CKKS FHE Parameter Configuration.}
\label{tab:hyperparameters}
\begin{tabular}{l|l|l|l}
\toprule
Category & Parameter & Symbol & Value \\
\midrule

\multirow{5}{*}{CKKS}
& Security Parameter & $\lambda$ & 80 \\
& Polynomial Degree & $N$ & $2^{18}$ \\
& Coefficient Modulus & $M$ & 100 \\
& Scaling Factor & $r$ & 30 \\
\midrule

\multirow{3}{*}{EWA}
& Plaintext Precision & $m$ & $10^{-4}$ \\
& Number of Companies & -- & 4000 \\
\midrule

\multirow{4}{*}{EMM}
& Plaintext Precision & $m$ & $10^{-4}$ \\
& Evaluation Categories & $n$ & 5 \\
& Credit Levels & $k$ & 6 \\
\midrule

\multirow{4}{*}{EDT}
& Maximum Tree Depth & $D$ & 8 \\
& Minimum Node Samples & $N_{\text{node}}$ & 10 \\
& Number of Features & $L$ & 8 \\

\bottomrule
\end{tabular}
\end{table}

\subsection{Performance Evaluation}

As shown in Table~\ref{tab:time_enc}, we list the parameters (except the FHE computation process) of a sampled real UAV-assisted secure wireless communication and computation system testing experiment. 
The results indicate that all operations in the prototype system are completed within seconds, which aligns with the expected performance 
of encrypted communication and computation. The overall latency remains 
well below the typical tens-of-seconds time scale required for flight operations, demonstrating satisfactory efficiency. 
A moderate time overhead is acceptable in exchange for enhanced privacy 
and security, effectively protecting the sensitive data of both parties 
during computation and communication.

\begin{table}[h]
\centering
\caption{Experiment Information of Sampled UAV-Assisted Secure Wireless Communication and Computation System.}


\label{tab:time_enc}
\begin{tabular}{l|c|l|c}
\toprule
\textbf{Parameter} & \textbf{Value} & \textbf{Parameter} & \textbf{Value} \\
\midrule
Key Generation & 0.43~s &
Transmission Distance & 5~m \\
Bluetooth Protocol & Bluetooth 6.0 & 
Transmission Time& 3.2~s \\
Encryption Time & 0.56~s &
Decryption Time & 0.02~s\\
\bottomrule
\end{tabular}
\end{table}

Table~\ref{tab:multi_node_results} shows that the EWA and EMM achieve high numerical precision, with relative errors (RE) on the order of $10^{-8}$ and $10^{-16}$, respectively. EWA maintains stable accuracy as the number of nodes increases, while EMM shows a negligible error increase with larger data scales. EDT consistently achieves prediction accuracy above 97\%, demonstrating the robustness of the proposed framework across varying computation data node lengths.
\begin{table}[h]
\centering
\caption{Performance Comparison under Different Numbers of Computation Data Nodes.}
\label{tab:multi_node_results}
\begin{tabular}{l|c|c|c}
\toprule
\textbf{Nodes} & \textbf{EWA RE} & \textbf{EMM RE} & \textbf{EDT Accuracy} \\
\midrule
100  & $2.21 \times 10^{-8}$  & $3.92 \times 10^{-17}$ & 98.1\% \\
200  & $1.13 \times 10^{-8}$  & $9.14 \times 10^{-17}$ & 98.2\% \\
500  & $1.76 \times 10^{-8}$  & $2.36 \times 10^{-16}$ & 98.4\% \\
1000 & $2.72 \times 10^{-8}$  & $5.79 \times 10^{-16}$ & 97.4\% \\
\bottomrule
\end{tabular}
\end{table}

The time performance of the proposed EWA, EDT, and EMM is shown in Table~\ref{tab:encrypted_time}. 
The computation time of three encrypted algorithms varies with the computation data node length. EWA exhibits excellent scalability, with time nearly constant at around 0.8~s. EMM grows linearly from 0.36~s to 3.48~s. EDT remains stable at approximately 5.4~s across all settings. 
The overall computation latency remains well below the tens-of-seconds time scale required for \textcolor{black}{coarse-grained} flight operations, demonstrating satisfactory efficiency. 
\begin{table}[h]
\centering
\caption{Encrypted Computation Running Time under Different Numbers of Computation Data Nodes.}
\label{tab:encrypted_time}
\begin{tabular}{l|l|l|l}
\toprule
\textbf{Nodes} & \textbf{EWA} & \textbf{EMM} & \textbf{EDT} \\
\midrule
100  & 0.81~s & 0.36~s & 5.42~s \\
200  & 0.82~s & 0.71~s & 5.41~s \\
500  & 0.81~s & 1.78~s & 5.50~s \\
1000 & 0.83~s & 3.48~s & 5.47~s \\
\bottomrule
\end{tabular}
\end{table}
The space overhead performance of the proposed EWA, EDT, and EMM is shown in Table~\ref{tab:encrypted_space}. 
EWA maintains a constant space overhead of $11.60$~MB regardless of the computation data node length. EMM shows linear growth from $39.31$~MB to $397.32$~MB as the FHE computation complexity increases with the matrix dimension. EDT remains stable around $450$ MB across all scales, as decision tree storage depends on model structure rather than data input. Considering the functional capabilities of the UAV onboard system, all space overhead is within acceptable limits.

\begin{table}[h]
\centering
\caption{Encrypted Space Overhead under Different Numbers of Computation Nodes.}
\label{tab:encrypted_space}
\begin{tabular}{l|c|c|c}
\toprule
\textbf{Nodes} & \textbf{EWA} & \textbf{EMM} & \textbf{EDT} \\
\midrule
100  & 11.60~MB & 39.31~MB & 441.26~MB \\
200  & 11.66~MB & 72.21~MB & 455.72~MB \\
500  & 11.73~MB & 164.45~MB & 468.75~MB \\
1000 & 11.88~MB & 397.32~MB & 494.39~MB \\
\bottomrule
\end{tabular}
\end{table}

Table~\ref{tab:matrix_results} compares the performance of our CKKS-based EMM with SHE and BGV-based methods at matrix dimension 100. These results demonstrate that CKKS offers superior time and space efficiency for encrypted matrix operations.
\begin{table}[h]
\centering
\caption{Performance Comparison with SHE and BGV Methods with the matrix dimension 100.}
\label{tab:matrix_results}
\begin{tabular}{l|c|c}
\toprule
\textbf{Algorithm} & \textbf{Running Time} & \textbf{Memory Consumption} \\
\midrule
CKKS-Based & \textbf{0.36~s} & \textbf{39.44~MB} \\
SHE-Based & 62.73~s & 162.96~MB \\
BGV-Based & 1.17~s & 87.25~MB \\
\bottomrule
\end{tabular}
\end{table}

Table~\ref{tab:dt_comparison} shows the comparison result between the proposed CKKS-based EDT and the BFV-based method. Results demonstrate that both algorithms yield \textcolor{black}{comparable accuracy.}
Although the proposed CKKS-based EDT incurs a larger space overhead, its time overhead is only one-third of that of the BFV method. In our experimental system, time efficiency is a more critical resource, as faster completion of encrypted computation tasks ensures that UAVs can finish the entire process during flight. Thus, the trade-off of increased space consumption for improved time performance is well justified,
\textcolor{black}{and the proposed CKKS-based EDT is more suitable for the considered UAV-assisted computing system.}
\textbf{Moreover, compared to the BFV, which only supports integer arithmetic, the CKKS also supports floating-point operations.}
\begin{table}[h]
\centering
\caption{Performance Comparison of Encrypted Decision Tree Algorithms with Computation Data Node Length 500.}
\label{tab:dt_comparison}
\begin{tabular}{lcc}
\toprule
\textbf{Metric} & \textbf{CKKS-Based EDT} & \textbf{BFV-Based EDT} \\
\midrule
Accuracy & 98.41\% & \textbf{98.82\%} \\
Running Time & \textbf{5.50~s} & 19.64~s \\
Memory Consumption & 468.75~MB & \textbf{207.21~MB} \\
\bottomrule
\end{tabular}
\end{table}
\section{Conclusion}
\label{conclusion}
In this paper, we propose a UAV-assisted privacy-preserving computing framework that integrates CKKS-based FHE with MPC for secure collaborative processing in wireless environments. Within this framework, we design three encrypted computation modules, EWA, EMM, and EDT, covering representative vector/matrix operations and machine learning tasks. Real-world benchmarks validate the feasibility of the proposed framework. Future work will focus on optimizing UAV trajectories, developing high-speed communication protocols, and extending the framework to support deep learning algorithms.

\bibliographystyle{unsrt}  
\bibliography{references}

\end{document}